\documentclass[10pt,conference]{IEEEtran}

\usepackage{graphicx}

\begin{document}

\title{Women Want to Learn Tech: \\ Lessons from the Czechitas Education Project}

\author{\IEEEauthorblockN{Barbora Buhnova$^{1,2}$ and Dita Prikrylova$^{1}$}
\IEEEauthorblockA{
$^{1}$\textit{Czechitas, Prague, Czech Republic} \\
$^{2}$\textit{Faculty of Informatics, Masaryk University, Brno, Czech Republic} \\
Email: \{baru,dita\}@czechitas.cz}
}

% The paper headers
\markboth{Journal of \LaTeX\ Class Files,~Vol.~14, No.~8, August~2015}%
{Shell \MakeLowercase{\textit{et al.}}: Bare Demo of IEEEtran.cls for IEEE Journals}
% The only time the second header will appear is for the odd numbered pages
% after the title page when using the twoside option.
% 
% *** Note that you probably will NOT want to include the author's ***
% *** name in the headers of peer review papers.                   ***
% You can use \ifCLASSOPTIONpeerreview for conditional compilation here if
% you desire.

% make the title area
\maketitle

\begin{abstract}
% Vime z rady studii, ze male holcicky ztraci zajem o IT
% Nicmene starsi divky maji druhou vlnu, kdy po IT pokukuji a je tezke pro ne sebrat odvahu a zkusit to
% In fakt, chteji se ucit, ale potrebuji helping hand a jiny pristuk k vyuce programovani
% Prezentujeme Czechitas, coz je XYZ

While it is understood by women that tech fluency might act as a powerful career accelerator or even a new career direction towards software engineering, this awakening often comes after graduation from a different field, when it is difficult for the women to make the shift towards tech and computing. In this paper, we report on our experience with running a successful education non-profit called Czechitas, which shows that women in their 20s and 30s are (maybe surprisingly) highly interested in learning tech, they just need a helping hand and tailored assistance, encouragement and guidance. 

\end{abstract}

% Note that keywords are not normally used for peerreview papers.
% \begin{IEEEkeywords}
%
% \end{IEEEkeywords}

% ------------------------------------------------------

\section{Introduction}

% Mini mitivacni gig ze v IT chybi lidi a ukazuje se, ze neni potreba aby tam bylo tak malo holek.
Across the world, as few as 10--25\% of tech professionals are women, with many women dropping their interest in tech for preventable reasons. This is a major issue for both the current and future jobs market, as Europe alone faces a shortage of hundreds of thousands skilled tech engineers to date \cite{Deloitte2016, Microsoft2017}. Moreover, the concerns are related also to the impact a tech background can have on a women's career, and the economic potential that accompanies it \cite{Forbes2012, Accenture2016equal}.
%"However, this problem also presents an opportunity. As Accenture noted in its earlier study, Getting to Equal — How Digital is Helping Close the Gender Gap at Work\cite{Accenture2016equal}, digital fluency acts as a powerful accelerant at every stage of a woman’s career." "But we have ample evidence that it is a key factor and acts as an accelerant in every stage of a person’s career — powerful in both education and employment, and increasingly important as women advance into the ranks of leadership." \cite{Accenture2016equal}

% Pokud se chceme bavit o tom, proc holky nejsou v IT, je potreba se podivat na to, v jakych cekovych intervalech je ztracime a v jakych cvilich se nam naopak chteji vratit ale my jim neumime jit naproti
% To better understand the phenomenon and find effective counter measures, the key is to understand what is the vulnerable age when girls loose interest in tech, and what on the other hand is the age when girls become open to regaining the interest. While study of the former period of life (let us call it \emph{vulnerable years}) can help us understand and mitigate the factors causing the drop in interest, the latter period (let us call it \emph{reconsideration years}) can help us to put effective measures in place that offer well targeted helping hand to these young women.

% Ztrate holek a duvodum se venuje treba studie Microsoftu, Accenture, zatimco Goole studie se snazi najit, co funguje, aby jsme make holky neztracely
Many studies have been conducted to date to reveal that in early education, girls and boys are equally touched by technology, computing and software engineering. But as adolescents (age 13 to 18), girls become much less interested in these topics, with limited recovery during the next 10 years that are crucial in terms of career choice \cite{Microsoft2017}. 

% Nicmene nikdo neresi tu vekovou skupinu kolem 25 let, kdy holky chteji do IT, ale chybi jim podana ruka
While these \emph{vulnerable years} and all factors that influence whether girls keep their interest in tech or not are very well studied and understood \cite{Microsoft2017, GoogleStudy2014, Accenture2016}, barely any attention is paid to the period in life of young women some years after graduation when they reconsider their career choice (we call it \emph{reconsideration years}, typically between 25 to 30 years of age). In our experience, this time might be linked to the moment when some of these women plan starting a family and consciously rethink what shall be the career they want to come back to after maternity leave. 

Experience shows that during these \emph{reconsideration years}, the interest in tech and software engineering among young woman raises again due to prospective opportunities that tech offers \cite{Microsoft2017, CzechitasPruzkum2017}. If however a helping hand is not in reach to the young women, the interest goes in vain.

\begin{figure}[htb]
    \centering
    \vspace{-2mm}
    \includegraphics[width=0.49\textwidth]{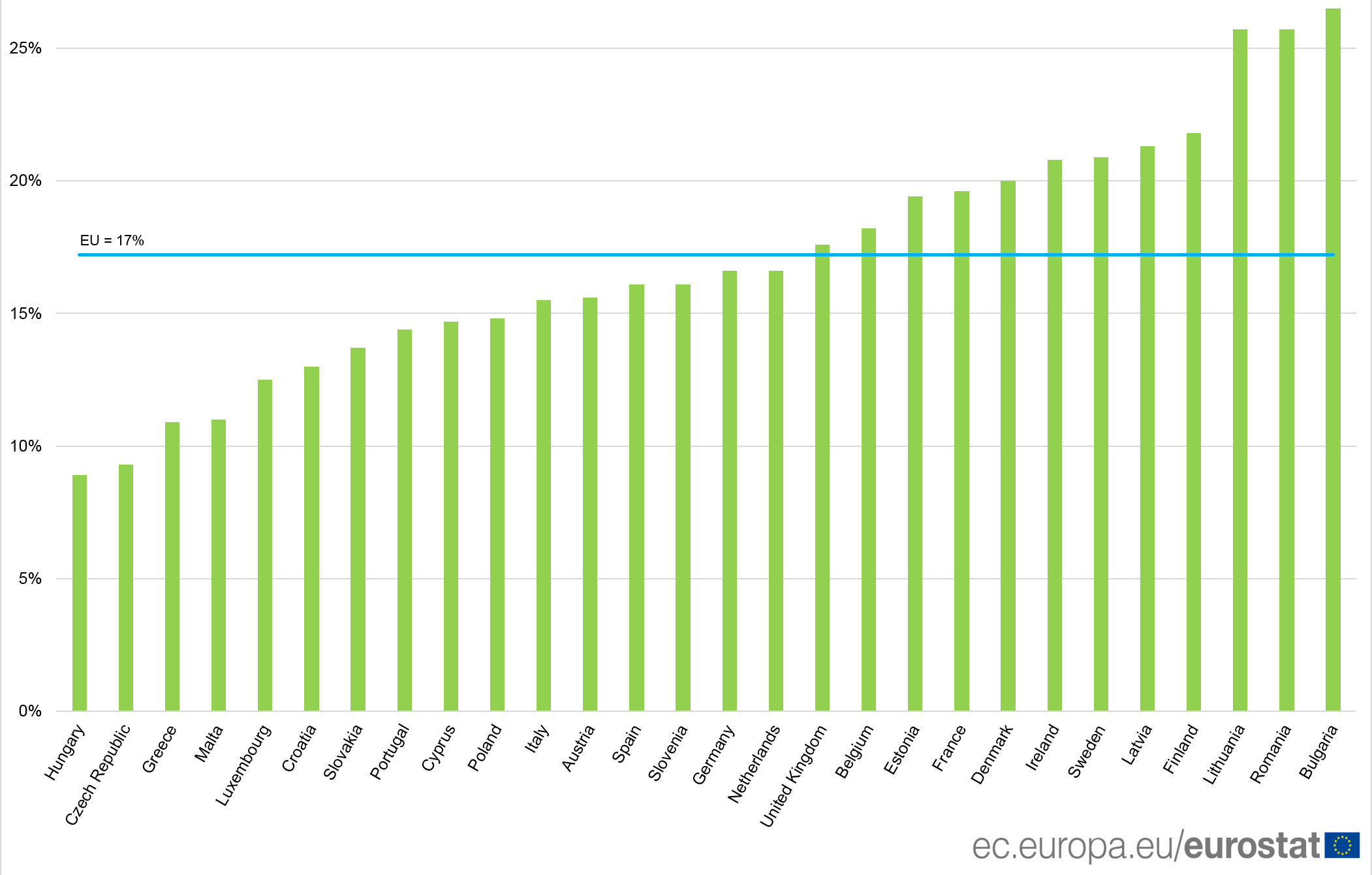}
    \vspace{-6mm}
    \caption{Women among ICT specialists in employment in 2017 \cite{Eurostat2017}}
    \label{fig:eurostat-2017}
    \vspace{-2mm}
\end{figure}

% Czechitas resi ty chvile
In this paper, we report on our experience with helping young women during these reconsideration years to find their way into tech, encourage and educate them. All that within the Czechitas non-profit education project, started specifically for this purpose in the Czech Republic, where the ratio of women in tech is the second worst in EU (see Figure \ref{fig:eurostat-2017}). Within Czechitas, we assist these young women in their learning path, we guide them towards specific technologies, and advise them in their career decisions. We connect them with professionals in the respective fields, and create welcoming community environment for them to learn from each other. Besides sharing this experience, this paper explains what kind of helping hand is needed during the reconsideration years so that the interest of these women in tech turns effective.

% ------------------------------------------------------

\section{The Story of Czechitas}
% Chceme holkam pomoct na zacatku, pomoct na stejnou startovni caru, pak uz to zvladnou mezi klukama

Czechitas \cite{CzechitasWeb, CzechitasRok2017} is a Czech non-profit organization that has emerged in 2014 from a simple idea to bring tech closer to girls, and girls closer to tech. Over the time, this idea has attracted a strong community of tech professionals, companies and volunteers, and gave rise to a portfolio of female-tailored courses in various areas of tech and software engineering, such as programming, web development, mobile app development, data science, testing, digital marketing and graphic design. 

\begin{figure}[bth]
    \centering
    \includegraphics[width=0.5\textwidth]{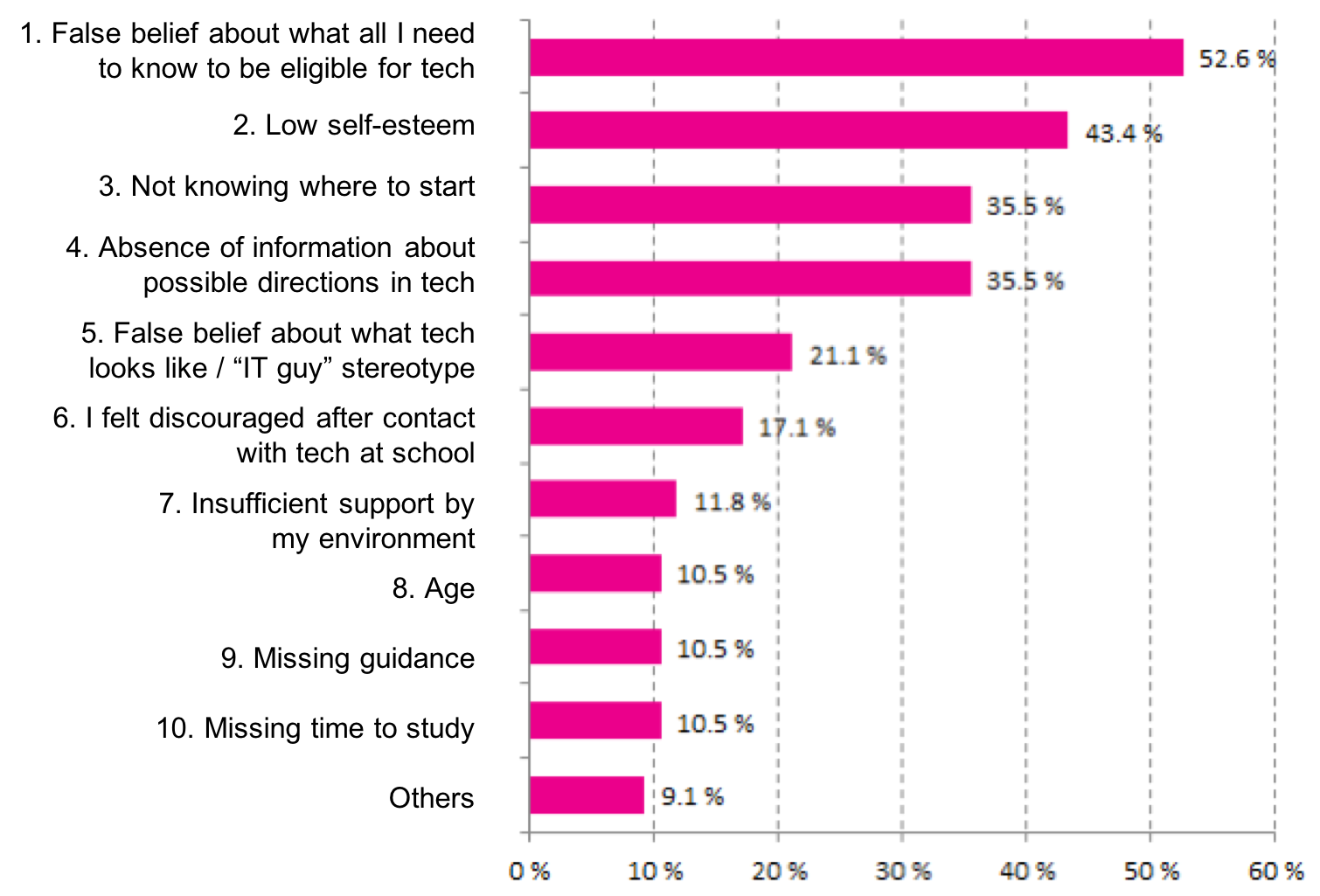}
    \vspace{-7mm}
    \caption{When looking back, what was the biggest obstacle for you when considering joining tech? \cite{CzechitasPruzkum2017}}
    \label{fig:czechitas-pruzkum-2017}
    \vspace{-3mm}
\end{figure}

Thanks to the success of our education activities, consisting of hundreds of events a year (each receiving multiple times more registrations than its capacity), we have become recognized as the leading platform in the Czech Republic actively addressing gender diversity in STEM. We have also been recognized abroad with numerous European as well as overseas awards, e.g. becoming the first organization from the Central Europe awarded the prestigious Google.org grant in 2015, and in 2018 becoming the first organization worldwide being awarded the Google.org grant for the second time \cite{CzechitasWeb}.

We have influenced over 10,000 women who graduated from our courses to either choose or change their career path to tech or use their new tech skills to advance their careers. Besides students, our community includes lecturers, professionals, partner companies, volunteers, the core team and social media audience (community of 15,000+). Even though the core team includes only 30 employees, we have the support of more than 350 volunteers actively involved in our activities (tech professionals, 40\% female).

% ------------------------------------------------------

\section{Key Factors when Reconsidering Tech}
% https://www.czechitas.cz/cs/blog/zeny-v-it/zeny-v-it-pruzkum-czechitas

Tech by itself is an appealing career choice, with high salary, job security, working flexibility, space for creativity and career growth. What is interesting then are the factors that negatively influence the decision to pursue tech career among women.

During the adolescence of girls, social encouragement and tech exposure in school, together with self perception and career perception are reported to be the key controllable indicators for whether or not girls decide to pursue a Computer Science or tech path \cite{GoogleStudy2014}. While numerous other studies exist that examine the \emph{vulnerable years} and confirm these findings \cite{Accenture2016, Microsoft2017}, little is known about the factors that play the key role during the \emph{reconsidearation years}, hindering the decision in favour of tech.

% \begin{enumerate}
% \item Social Encouragement: Positive reinforcement of Computer Science pursuits from family and peers.
% \item Self Perception: An interest in puzzles and problem solving and a belief that those skills can be translated to a successful career.
% \item Academic Exposure: The availability of, and opportunity to participate in, structured (e.g., graded studies) and unstructured (e.g., after-school programs) Computer Science coursework.
% \item Career Perception: The familiarity with, and perception of, Computer Science as a career with diverse applications and a broad potential for positive societal impact.
% \end{enumerate}

In 2017 we have surveyed 302 participants of Czechitas events together with representatives of 36 companies to understand these factors \cite{CzechitasPruzkum2017}. Our findings (see Figure \ref{fig:czechitas-pruzkum-2017}) very well match the conclusions of Harvey Mudd's President, Maria Klawe, who compiled their experience in three points, saying "Number one is they think it's not interesting. Number two, they think they wouldn't be good at it. Number three, they think they will be working with a number of people that they just wouldn't feel comfortable or happy working alongside \cite{Forbes2012}." Let us elaborate on these three points and add our observations, resulting also in a fourth point, which is that young women during reconsideration years critically miss guidance in learning.

\subsection{They think it's not interesting}

\begin{figure}[htb]
    \centering
    \includegraphics[width=0.4\textwidth]{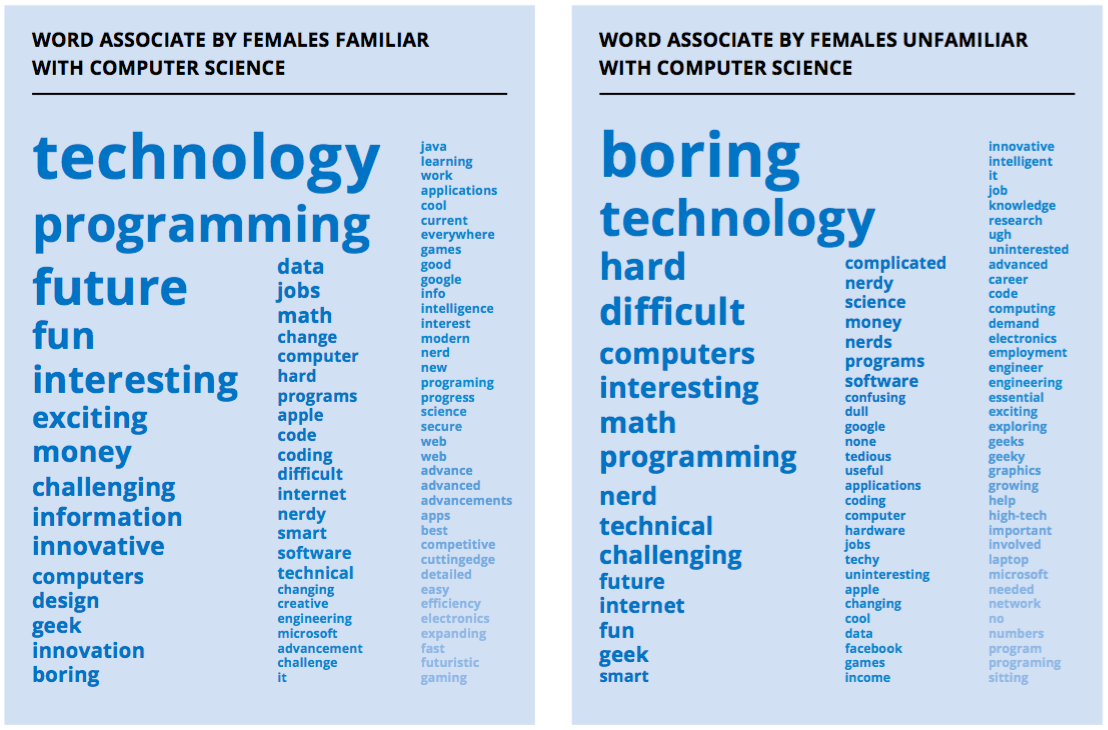}
    \vspace{-3mm}
    \caption{Word associate by females (un)familiar with computer science \cite{GoogleStudy2014}}
    \label{fig:google-words-2014}
    \vspace{-3mm}
\end{figure}

Alarming effect of low exposure of girls to tech was described by Google in 2014, showing that the unfamiliarity with tech alone causes inclination to negative connotations with tech as a domain \cite{GoogleStudy2014}. The study found that no matter if the girls were exposed to tech via compulsory or elective classes, the exposure itself was enough for them to characterize tech in much more positive terms. While girls unfamiliar with tech were using words like "boring, hard, difficult", girls previously exposed to tech used words like "future, fun, interesting, exciting" (see Figure \ref{fig:google-words-2014}).

One of the first corrective tasks we need to address is then to turn around the perception of tech that young women have due to their low exposure to tech in the past. Without that, young women have difficulty to perceive tech as a career that fulfills both the creative passion (inventing, problem solving, exploration, etc.) and the intangible, social passions (helping people, conservation, medical breakthroughs, etc.) that make a profession personally rewarding \cite{GoogleStudy2014}.

\subsection{They think they wouldn't be good at it}

The key factor in our study among women participants of Czechitas courses is the perception of their own proficiency in tech (see Figure \ref{fig:czechitas-pruzkum-2017}) \cite{CzechitasPruzkum2017, GoogleStudy2014}. Interestingly, the perception is rooted more strongly in their confidence than objective reasons, which is commonly observed as the Confidence Gap between men and women reported in various fields \cite{ConfidenceGap2014}.

Similar conclusions have been drawn in mathematics, where although standardized tests indicate that girls score just as well as boys in math, and that in school grades, girls often even outperform boys \cite{Science2008}, girls tend to have higher levels of math anxiety and lower levels of confidence in their math skills. In other words, even when girls show similar performance levels to boys, they are often less sure of themselves, which applies to math as well as tech \cite{GirlsMath2018}.

\subsection{Who they will be working with}

In our experience, women we teach are highly concerned about the work environment they shall once belong to, being afraid that they might end up working with people that they would not feel comfortable working alongside \cite{Forbes2012}. And indeed, some studies report that 27\% of women cited discomfort with their work environment as a factor why they left their IT job. Furthermore, workplace policies not suited to women also play a role \cite{Deloitte2016}. Our study moreover shows that women perceive work in tech jobs as absent of communication among people (only 1\% considered soft skills useful in tech career in our study \cite{CzechitasPruzkum2017}), which they do not feel comfortable with.

\subsection{Missing guidance}
\vspace{-1mm}

In one form or another, young women during their reconsideration years feel lost when navigating tech education on their own. Helping hand or guidance during that time is at the end what decides whether they engage in tech education or not. Sometimes, they only need a direction, such as somebody who helps them to decide what programming language to start with. There is no universal answer to this question, which is why they often feel confused when they try to find the answer online. Moreover, they often ask how much time they shall be ready to invest before seeing results (e.g. in terms of an application they would like to develop), what learning path takes them to a specific job interview, which of the thousands of online courses they shall start studying. Then during learning, they often get stuck on trivial mistakes, which they find impossible to debug without help of another person. And they miss community of like minded female students.

% ------------------------------------------------------

\section{Our Solution}
\vspace{-1mm}

Research by Accenture and Girls Who Code concludes that improving universal access of women and girls to tech and computing will not by itself address the gender gap. They instead suggest that only by tailoring courses to girls’ specific needs can we boost their commitment to computing \cite{Accenture2016}. In this section, we discuss what such tailoring means to us.

\subsection{Community}
\vspace{-1mm}
% Girls miss the community they would belong to
% We create such community for them
% Also offline events enforce community learning
% Food sharing
% Informal atmosphere
% It made the concept very popular

We build a community of girls and women who learn to code, as we see that community feeling is one of the key factors in retaining these tech students. All our events have very informal and friendly atmosphere, with team building activities and food sharing, to enforce the community feeling.

\subsection{Safe environment and encouragement}
\vspace{-1mm}
% The core are saturday learnings
% Safe to make mistakes
% Individual pace - necessary to let everybody to feel safe to make mistake and experiment the way they need it
% Individual mentors
% Bonus excercises
% Variety of activities to make it more fun

Besides a lecturer, each of our coding events (for 30 participants) features around 5 more teacher mentors (we call them couches) who provide individualized help, support and encouragement to the students. Couches often sit by the students, helping them with whatever difficulty they have, which creates friendly atmosphere in which the students feel comfortable asking questions and reporting troubles they are facing (i.e. safe environment for trial and error). It moreover facilitates varying pace of each participant, thanks to which everybody can study on the speed that best fits their needs. This is in line with a study that shows that over a half of the girls learning tech would like to receive more active encouragement from teachers \cite{Microsoft2017}.

\subsection{Practical experience and hands-on exercises}
\vspace{-1mm}
% Hackathons

All our events feature many fun and creative hands-on exercises. Studies show that the more practical experience a girl receives during her education--–inside or outside the classroom–--the higher interest in STEM she builds \cite{Microsoft2017}. Creativity in the classroom is also key. Girls who like STEM enthuse about being able to choose their own projects or go on field trips where STEM subjects are brought to life.

\subsection{Real-life applications}
\vspace{-1mm}
% Coding clubs, internships, mentoring program

Our experience supported by studies also shows that women become more interested in tech once they are able to conceive how tech can be applied to real-life situations and how relevant it might be to their future \cite{Microsoft2017}. This is why we integrate real-life applications and projects in our activities.

\subsection{Role models}
\vspace{-1mm}
% Meetups, but not only with top experts, but also fresh graduates from our courses
% Graduates as couches
% Cena Czechitas

Having visible female role models sparks women interest in tech careers and helps them to picture themselves pursuing these fields. This is why we organize meet-ups with inspirational female engineers and organize Czechitas Thesis Award, which showcases examples of excellent graduation theses in tech from Czech universities, drawing attention to talented girls who are passionate about tech.

\subsection{Career guidance}
\vspace{-1mm}
% 75% uvazuje, ze by se jednou IT chtely zivit, ale nase kurzy jsou jen zacatek
% Internships
% DA
% Mentors
% Job fair

We offer women a helping hand on their career path, from the initial decision about the field of study to the job interview. We organize annual Job Fair, which is a vibrant event where our community of students, mentors, lecturers and volunteers can meet our partner companies. The vision of this event is to get these two groups together for one day and provide them with specialized content, including career and IT workshops for our community and gender or educational workshops for our partner companies.

\subsection{Partnering with companies}
\vspace{-1mm}
% Show girls the environment
% It’s important that they see what IT is really about
% Cite the google report
% Educate the companies - gender bias within the hiring process, be more sensitive

It has been recognized that young women are more likely to pursue tech careers when they are confident that men and women will be treated equally working in these disciplines \cite{Microsoft2017}. We have observed that women feel much more confident when filing their job application to a Czechitas partner company, because the fact that the company partners with us gives them the reassurance that they are welcome at job interview. We also educate our partner companies about gender bias within the hiring process and what makes them more/less attractive employer to women.

% ------------------------------------------------------

\section{Format of Events}
\vspace{-1mm}
\label{sec:format}

% Full saturday
% Accessible to mothers
% Selection based on motivation
% Temata
% Co je nejoblibenejsi
% I online na chci se ucit

\subsection{One-day workshops}
\vspace{-1mm}

The most popular format of our events are full-day workshops, taking place typically on Saturdays or Sundays. These focus on introduction to programming, web and mobile apps development, embedded systems, graphic design, data science and digital marketing. Most popular programming languages are Javascript, Python, Java, C\#, PHP, Kotlin, and Swift. The topics also include various software engineering essentials, such as testing, agile development, project management, security engineering, or version control.

The aim of one-day workshops is to spark interest in the attendees about coding and other areas of software engineering, and to give them necessary encouragement and guidance to study further.

\subsection{Long-term courses}
\vspace{-1mm}

We yearly organize over 20 spring and autumn courses, focusing mainly on various programming languages and technologies (most popular are Java, C\#, Javascript, PHP, HTML/CSS, Android, UX design, testing). Each of these courses comprises of 10-14 evening sessions of 2-3 hours each. The aim of long-term courses is to teach women a specific programming language or tech skill, and equip them with confidence for starting their own software development.

\subsection{Re-qualification academies}
\vspace{-1mm}

For the women who made the decision to change the career, we offer an intense re-qualification program, which involves three months of lectures, hands-on exercises, and mentoring, called Digital Academy. Throughout the Academy, students participate in internships with companies and work on their final graduation projects. We offer them career services focused on professional transit to tech.
 
% We reserve 20\% of the Academy seats for mothers with small children, helping them to restart their career after maternity leave, which on average 6 years long in the Czech Republic. Another 20\% of the seats are reserved for fresh graduates who have difficulties to find work in their field of study.

We have developed this unique type of re-qualification program (with successful 70\% placement of the graduates on junior job positions in partner companies) for two fields: (1) data science and (2) backend web development in Java. The \emph{Digital Academy: Data} includes teaching blocks on data mining, data processing and visualization tools, statistics, digital marketing, SQL and databases, data science and discovery, and Python as the programming language. The \emph{Digital Academy: Java} focuses on key skills and technologies needed to develop web application with the focus on its backend in Java.

\subsection{Hackathons}
\vspace{-1mm}

To promote creativity and confidence among participants of our courses, we organize hackathons, where the women can test their knowledge during intense work on a full frame team project under the supervision of experienced mentors. The hackathons are organized both as stand-alone events and as part of the academies and other intense education programs.

\subsection{Coding clubs and Internships}
\vspace{-1mm}

Coding clubs and internships give girls the opportunity to work on a long-term team coding project under the supervision of an experienced professional. In case of the coding club, the team meets 2-4 times per month to discuss their progress on the joint project. In case of the internship, the woman becomes part of a team in a company that provides support for her to learn in practice.

% \subsection{Job fairs}
% 
% Czechitas Job Fair is a vibrant event where our community, including students, mentors, lecturers and volunteers can meet our partner companies. The vision of this event is to get these two groups together for one day and provide them with specialized content, including career and IT workshops for our community and gender or educational workshops for our partner companies. An inseparable part of the Job Fair is an interactive program for both groups together, such as one-minute networking. During the entire day, there is a daycare service for children with educational activities, so that the students can focus on their careers.

% \subsection{Czechitas Thesis Award}
% 
% The idea of the Czechitas Thesis Award is to showcase examples of excellent graduation theses in tech from Czech universities, as they set a great example to follow by girls deciding about their future field of study. The project comprises of a mentoring program with our partner companies and a financial award for the winners. The award aims at drawing attention to talented girls who are passionate about tech and helping them implement their projects with the support of experts from their respective fields. 

\subsection{Czechtias New Generation}
\vspace{-1mm}
% Czechitas nova generace
% Letni skola

Over the years, we have become recognized as a platform with the ability to guide newcomers to tech. Thanks to that, we have been invited to engage in tech education of children and youth in the Czech Republic. Since then, we have not only been active in advancing tech education among kids (engaging female role models in teaching), but also in equipping their school teachers with the necessary skills to make tech education accessible and fun. 

% Apart from one-day workshops, we focus mainly on intensive forms of education, such as summer camps or weekend camps. These activities vary in their target group and topics. We teach kids from 8 years of age, as well as focus on high-school girls or boys. We emphasize influencing and motivating younger girls and female students to introduce them to the field of IT and perhaps aid them in their pursuit of career choices. 

% We organize seminars for teachers to enhance and refine their teaching lessons. Apart from these trainings of technical skills, we offer an organizational platform for sharing knowledge and methodical support on their way. We put special emphasis on activation of the school headmasters who have the real power to influence the way of education in their schools. 

% ------------------------------------------------------

\section{Conclusion}

This paper reports on our successful education project, called Czechitas, which supports women when restarting their career and changing direction towards tech and computing. The project enjoys large popularity, and hence is now growing into multiple new cities within the Czech Republic (from 5 cities where we have our teams now, growing to 3 more cities in the next two years), while enlarging its portfolio with new topics and forms of learning (building an online learning platform now), as well as support for educators (training of tech teaching skills, organizational platform for knowledge sharing and methodical support). Moreover, we are now putting effort in improving the way Czech schools approach tech education, to prevent the low female engagement in tech, which we are correcting now.

% ------------------------------------------------------

\bibliographystyle{IEEEtran}

\end{document}